\documentclass[%
 reprint,
 amsmath,amssymb,longbibliography,
 aps
]{revtex4-1}

\usepackage{graphicx}
\usepackage{dcolumn}
\usepackage{bm}

\usepackage{braket}

\def\bcen{\begin{center}}
\def\ecen{\end{center}}
\usepackage{epstopdf}
\usepackage{soul}
\usepackage{epstopdf}
\usepackage[breaklinks=true,colorlinks]{hyperref}
\begin{document}

\preprint{APS/123-QED}

\title{Limitations to Realize Quantum Zeno Effect in Beam Splitter Array - a Monte Carlo Wavefunction Analysis}

\author{Nilakantha Meher$^1$}
\email{nilakantha.meher6@gmail.com}
\author{Akhil Raman$^2$}
\author{S. Sivakumar$^3$}%
 \email{sivakumar.srinivasan@krea.edu.in}
\affiliation{%
 $^1$Department of Chemical and Biological Physics,
Weizmann Institute of Science, Rehovot 7610001, Israel\\
$^3$School of Physics, University of Hyderabad, Andhra Pradesh 500046, India\\
$^3$Division of Sciences, Krea University, Andhra Pradesh 517646, India
}%




\date{\today}
\begin{abstract}
Effects of non-ideal optical components in realizing quantum Zeno effect in an all-optical setup are analyzed.  Beam splitters are the important components in this experimental configuration.  
Non-uniform transmission coefficient, photon absorption and  thermal noise  are considered. Numerical simulation of the experiment is performed using the Monte Carlo wavefunction method.    It is argued that there is an optimal number of beam splitters to be used for maximizing the expected output in the experiment.
\end{abstract}

\pacs{Valid PACS appear here}
\maketitle

\section{Introduction}
 A closed quantum system's evolution is governed by the Schrodinger equation.  If a system starts in an eigenstate of the Hamiltonian, it remains in that state subsequently if it evolves under the Hamiltonian. If the initial state is not an eigenstate or if there is an external interaction, then the system will evolve in a non-trivial way.  However,   irrespective of the initial state of the system,  it is possible to slow down or stall the evolution of the system from its initial state. It is achieved through repeated measurements to determine if the system is in its initial state or not.  Quantum Zeno effect (QZE) refers to the possibility of using repeated  measurements  to stall the evolution of the state [1,2].  This situation is very akin to the argument by Zeno that motion (evolution) is impossible, and hence the name quantum Zeno effect [1,3-5]. Briefly, in the absence of dissipation or decoherence, evolution of a quantum system is given by 
$\ket{\psi(t)}=e^{-iHt}\ket{\psi(0)},$
where $H$ is the Hamiltonian of the system, $\ket{\psi(0)}$ is the initial state and $\ket{\psi(t)}$ is the state at time '$t$'. Consider a measurement to determine if the system is in its initial state.  In this measurement, the probability of finding the system  in its initial state $\ket{\psi(0)}$ after a small interval of time $\delta t$ is
$P(\delta t)=|\langle \psi(0)|\psi(\delta t)\rangle|^2\approx 1-(\delta t)^2(\Delta H)^2,$ 
where $ \Delta H=\sqrt{\bra{\psi(0)} H^2\ket{\psi(0)}-\bra{\psi(0)} H\ket{\psi(0)}^2}$.   Higher order terms in $\delta t$ have been neglected as $\delta t $ is  small. If $N$ such measurements are made in a time interval $T$, the probability of detecting the system in $\ket{\psi(0)}$ after all the measurements are completed  is 
$P_N(T)\approx\left(1-\frac{T^2}{N^2}( \Delta H)^2\right)^N\approx \left(\cos^2\left(\frac{T}{N}\Delta H\right)\right)^N,$ 
where $T/N$ is an infinitesimal time interval [6].
In the limit of $N \rightarrow \infty$, the probability of finding the system in its initial state is unity for all time. Essentially, continuous measurement inhibits the evolution of the system, which is the quantum Zeno effect [1].

\begin{figure}[h!]
\centering
\includegraphics[height=3cm,width=9cm]{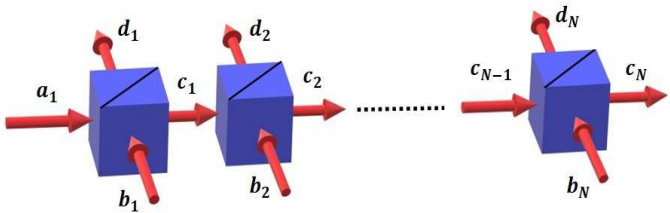}
\caption{(Color online) Beam splitter array to demonstrate QZE.   The photon output from the transmission port of each beam splitter is the input to the next one. }
\label{setup}
\end{figure}

Realization of QZE has been discussed in  different physical systems [7-18]. Agarwal and Tewari proposed an  interesting and easily implementable scheme of QZE in an array of $N$ identical beam splitters (Fig. \ref{setup}) [19]. The interest in QZE is not because of its fundamental significance alone, but also due to potential applications such as counterfactual communication between two parties [20], wherein information is transmitted without a transfer of  any physical particle. In the asymptotic limit of large $N$, which is a requirement for QZE in the considered setup, the two parties can communicate with an arbitrarily large efficiency under ideal conditions. However, the experimental imperfections such as non-uniform transmission coefficients, photon absorption, thermal noise, etc. in the beam splitters affect the efficiency of communication. Hence, it is of interest to study the role of such imperfections on realizing QZE in an array of beam splitters. The analysis is presented here in the context of an array of beam splitters (Fig. \ref{setup}) in which a single photon undergoes either reflection or transmission in each beam splitter. Similar analysis can be done for a few other experimental configurations such as the inhibition of polarization rotation of photons passing through an optical medium [21] and the transfer of a photon through an array of coupled cavities.  The reason that the analysis applies to these systems is that the  unitary transformation relevant to all these configurations is of the form  $e^{-iJ(a^\dagger b+a b^\dagger)}$ (equivalent to beam splitter transformation), where the operators are suitably interpreted to mean creation and annihilation of polarization modes or photons in cavity modes.   

To realize QZE, the arrangement is such that one of the outputs of a beam splitter is the input to the next beam splitter (Fig. \ref{setup}). For instance, output from the port $c_j$ of $j$th beam splitter is the input to  $(j+1)$th beam splitter. However, the output ports $\{d_j\}, (j=1,2,3\cdots N)$ are kept open in the sense no measurement is carried out in  these ports. This is to introduce a loss of coherence at the output state to mimic occurrence of  decoherence upon measurement.
Now, a single photon is input into the port $a_1$ while  the other input ports $\{b_j\}$ have their corresponding vacuua as input.   Then the state after the first beam splitter is
$\ket{\psi_{c_1d_1}}=\cos\theta\ket{10}-\sin\theta\ket{01},$
where $\cos\theta$ is the transmission amplitude and $\sin\theta$ is the reflection amplitude [22,23]. The state $\ket{10}$ represents  a single photon transmitted along the port $c_1$ and no photon along $d_1$. This can be seen more directly by  calculating the state of the field along the two output ports of the beam splitter.  The input-output relations relating the amplitudes of the input and output states are $a_1=c_1\cos\theta-d_1 \sin\theta$ and $b_1=c_1\sin\theta+d_1 \cos\theta$.  The state of the field at the output port $c_1$ is calculated by tracing over the port $d_1$ to get $
\rho_{{c_1}}=\text{Tr}_{d_1}(\ket{\psi_{c_1d_1}}\bra{\psi_{c_1d_1}})=\cos^2\theta \ket{1}+\sin^2\theta\ket{0}.$ 
Thus, the probability of finding the photon at the port $c_1$ is $\cos^2\theta$. The state $\rho_{c1}$ is the input to the second beam splitter.  Similar calculations can be done to calculate the input and output fields of the successive beam splitters.  If the photon input into the first beam splitter has to exit the array, it has to be transmitted by each of the beam splitters.  These being independent events, the  probability of detecting the photon at the end of the array is the product of the probabilities for transmitting a photon by each of the beam splitters. Therefore,  the net probability $P_{1}(N)$ to receive a photon at the port $c_N$ is 
\begin{align}\label{QZEbeamsplitter}
P_{1}(N)=\cos^{2N}\theta.
\end{align} 
The subscript "1" is to imply that the expression is relevant to single-photon transmission.  
If $\theta=\frac{\pi}{2N}$, the single photon detection probability $P_{1}(N)$ is  unity in the limit  $N\rightarrow \infty$, which is analogous to the quantum Zeno effect [19]. It is to be noted that the parameters $\theta$ and $N$ are related through $\theta=\pi/2N$. Hence, to maximize the probability $P_1(N)$, one has to increase the number of beam splitters in the array with reduced values of $\theta$ such that $\theta=\pi/2N$. 
\section{QZE in imperfect beam splitters}
To demonstrate QZE with  beam splitter array,  $\theta$ is chosen to be $\pi/2N$, where $N$ is the number of beam splitters in the array. However, in practical situations, realizing precise values of $\theta$ is difficult [24]. Any defect in the beam splitter will change the probability of transmission [25]. In this section, we discuss QZE if  beam splitters used in the array are not identical, that is, the value of $\theta$ is not exactly $\pi/2N$ for all the beam splitters.  However, the average of $\theta$ of all the beam splitters is $\pi/2N$. Hence, for $j$th beam splitter, the parameter $\theta_j$ is chosen randomly from a normal distribution with mean $\bar{\theta}=\pi/2N$ and standard deviation $\sigma$. Larger $\sigma$ indicates larger range of possible values of $\theta$ in the beam splitter. Then, we calculate the single photon detection probability $P_1(N)$ at the end of the array as an ensemble average the probabilities realized for various choices of ${\theta_j}$.

\begin{figure}
\centering
\includegraphics[height=6cm,width=9cm]{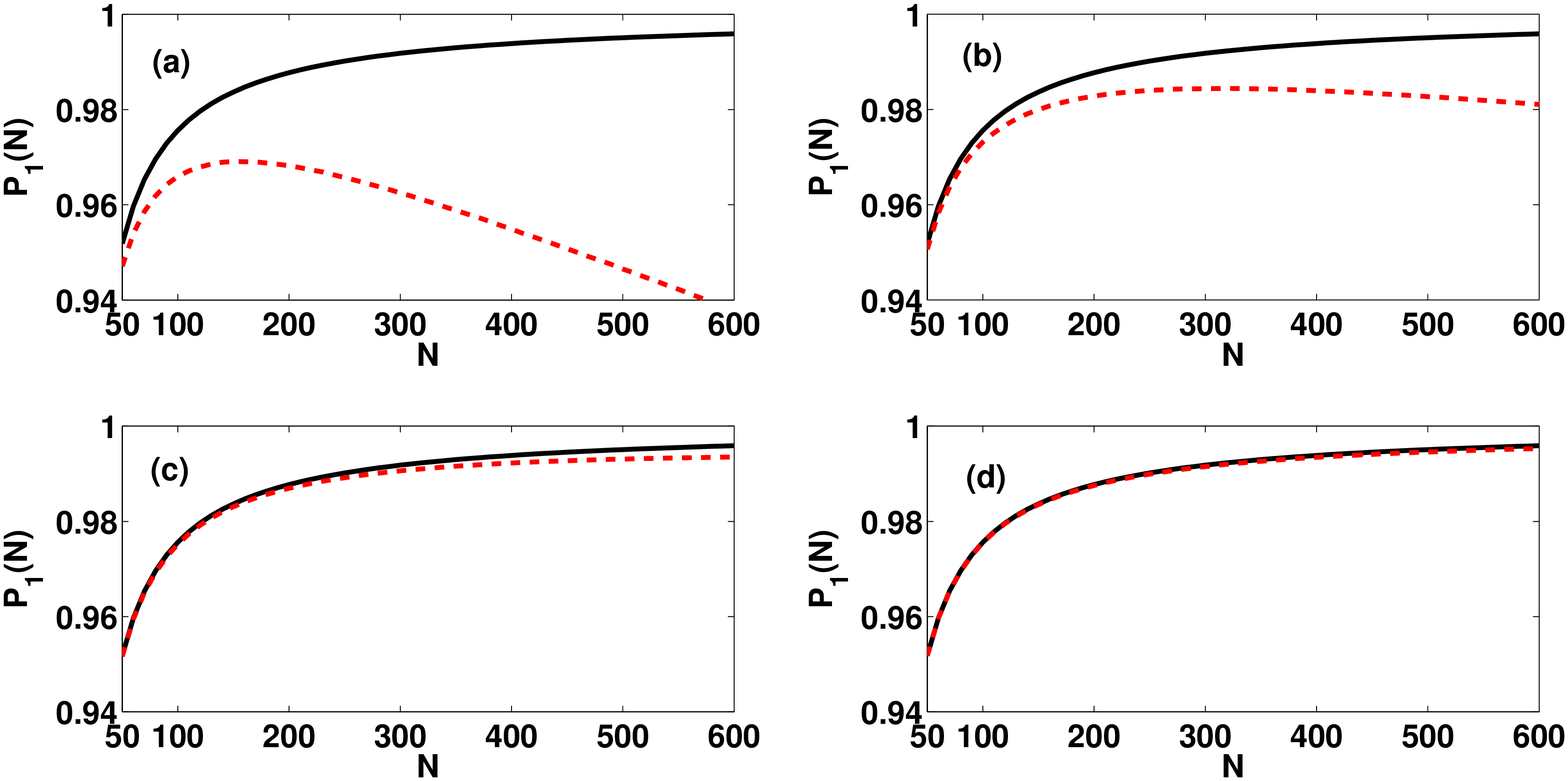}
\caption{(Color online) Single photon detection probability at the end port in the arrangement consisting of $N$ number of beam splitters. The parameter $\theta$ is not the same for all the beam splitters.  The parameter $\theta$ is normally distributed with mean $\pi/2N$ and standard deviation $(a)\sigma=1/100, (b)\sigma=1/200, (c)\sigma=1/500, (d)\sigma=1/1000$. Dashed line corresponds to the non-ideal case (Eqn. \ref{imperfectZeno}) which is compared with ideal case Eqn. \ref{QZEbeamsplitter} (continuous line).}
\label{imperfection}
\end{figure}

Let the set $\{\theta_1, \theta_2,....,\theta_N\}$ is sampled from a normal distribution, where $\theta_j$ corresponds to $j$th beam splitter in the array. If a single photon is input to the port $a_1$ and the vacuum as the input to all the ports $\{b_j\}$, then the state at the end of the array, that is, at the port $c_N$ (Fig. 1) is 
\begin{align}
\rho_{c_N}&=\cos^2\theta_1 \cos^2\theta_2...\cos^2\theta_N\ket{1}\bra{1}\nonumber\\
&+[\sin^2\theta_1+\cos^2\theta_1\sin^2\theta_2+\cos^2\theta_1\cos^2\theta_2\sin^2\theta_3+...\nonumber\\
          &...+\cos^2\theta_1\cos^2\theta_2...\cos^2\theta_{N-1}\sin^2\theta_N]\ket{0}\bra{0}.  
\end{align} 
The probability of detecting the single photon at the port $c_N$ is
\begin{align}\label{imperfectZeno}
P_1(N)=\cos^2\theta_1\cos^2\theta_2...\cos^2\theta_N.
\end{align}
It is to be noted that if the beam splitters are identical, that is, $\theta_1=\theta_2=...\theta_N=\theta$, we recover Eqn. \ref{QZEbeamsplitter}. 
The variation of $P_1(N)$ with respect to $N$ is shown in Fig. \ref{imperfection}, for different values of the standard deviation $\sigma$ in the parameter $\theta$.      
The single photon detection probability $P_1(N)$ does not become unity with increasing $N$ for large $\sigma$.  This feature is shown in Fig. \ref{imperfection} (dashed line) along with the ideal case situation when every beam splitter has the same value of $\theta$.  At smaller values of $\sigma$, $P_1(N)$ approaches unity as $N$ increases.     Hence, nearly identical beam splitters are required to realize QZE.   This is a drawback of the configuration shown in Fig. \ref{setup}.    

\begin{figure}
\centering
\includegraphics[height=5.4cm,width=8.5cm]{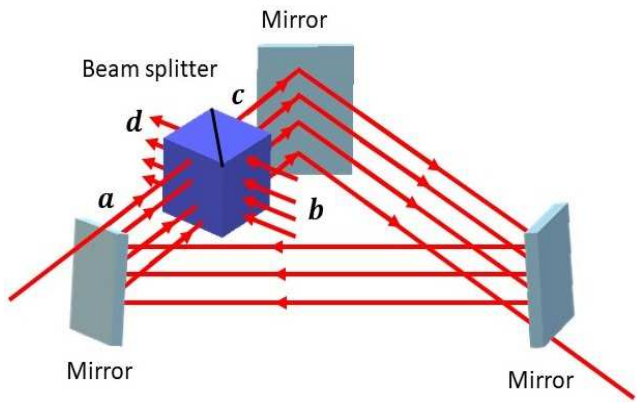}
\caption{(Color online) An alternate configuration to the array configuration shown  in Fig. \ref{setup}.  This arrangement involves only one beam splitter.  This helps to  overcome the stringent requirement for identical  beam splitters for the array configuration.} 
\label{ThreeMirrors}
\end{figure}

The problem of unequal values of $\theta$  can be overcome in the configuration  shown in Fig. \ref{ThreeMirrors}. The setup consists of a beam splitter and three mirrors. Properly aligning the three mirrors makes the input field traverse through the beam splitter many times.  The number of traversals is equal to the number of identical beam splitters in the array. The ports $a$ and $b$ are the input ports whereas $c$ and $d$  are the output ports of the beam splitter. A similar configuration having three mirrors and a polarization rotator has been used to demonstrate QZE  on the polarization degree of freedom of a photon [26].
\section{QZE in the presence of thermal noise} 
In this section, the effect of input thermal field, instead of the vacuum,  at the input ports $\{b_j\}$ on QZE is discussed.   To delineate the effect of the thermal noise,  $\{\theta_j\}$ is taken to be uniform. The vacuum state in the input port $\{b_j\}$ is meaningful if the temperature of the beam splitters is absolute zero.  At non-zero temperatures, there will always be thermal field. Hence, thermal states with small average number of photons (low temperature) are considered  as the input states.  At low temperatures, the thermal field is primarily the vacuum state while the higher number states make much smaller and monotonically decreasing contributions [23]. In this approximation,  the respective input states at all the ports $\{b_j\}$ are 
\begin{align}
\rho_{th}\approx\alpha\ket{0}\bra{0}+\beta \ket{1}\bra{1}.
\end{align} 
Here $\alpha \propto \frac{1}{(1+\bar{n})}$ and $\beta\propto \frac{\bar{n}}{(1+\bar{n})^2}$, where $\bar{n}$ is the average number of photons in the thermal field [23].   As $\bar{n}<<1$, $\alpha>>\beta$ and thus, the contributions from higher order photon number states are neglected.

Now, a single photon is made to pass through the input port $a_1$ and the thermal field $\rho_{th}$ at the port $b_1$, then the state after the first beam splitter is [27]
\begin{align}
\rho_{c_1d_1}=\alpha \ket{\phi_1}\bra{\phi_1}+\beta \ket{\phi_2}\bra{\phi_2},
\end{align}
where $\ket{\phi_1}=\cos\theta \ket{10}-\sin\theta \ket{01}$ and $\ket{\phi_2}=\frac{1}{\sqrt{2}}(\sin2\theta\ket{20}+\sqrt{2}\cos2\theta\ket{11}-\sin2\theta \ket{02})$. As we are interested in the state at the transmission port $c_1$, we take trace over the reflection port $d_1$ and get the state $\rho_{c_1}$ to be
\begin{align}
\rho_{c_1}=\text{Tr}_{d_1}(\rho_{c_1d_1})=A_0\ket{0}\bra{0}+A_1\ket{1}\bra{1}+A_2\ket{2}\bra{2},
\end{align}
where $A_0=\alpha \sin^2\theta+2\beta\cos^2\theta\sin^2\theta$, $A_1=\alpha\cos^2\theta+\beta\cos^2 2\theta$ and $A_2=2\beta\cos^2\theta\sin^2\theta$.
The probability of detecting the single photon in $\rho_{c_1}$ is 
$P_1(1)=A_1=\alpha\cos^2\theta+\beta\cos^2 2\theta \approx \alpha\cos^2\theta,$
for $\beta<<\alpha$ and small $\theta$. Now, if the state $\rho_{c_1}$ is input to the second beam splitter, then the state at the port $c_2$ will be
\begin{align}
\rho_{c_2}=&(A_0\alpha+A_0\beta \cos^2 \theta +A_1 \alpha \sin^2\theta+..)\ket{0}\bra{0}\nonumber\\
&+(A_1^2+ A_0\beta \sin^2 \theta+A_2\alpha \sin^2 2\theta)\ket{1}\bra{1} \nonumber\\
+&(\tfrac{1}{2}A_1\beta \sin^2 2\theta+2A_2 \alpha  \cos^4 \theta+..)\ket{2}\bra{2},  
\end{align}
and the probability of finding a single photon is
$P_1(2)=A_1^2+ A_0\beta \sin^2 \theta+A_2\alpha \sin^2 2\theta \approx \alpha^2 \cos^4 \theta.$ 
Proceeding with a similar manner, the probability of detecting a single photon at the port $c_N$ will be
\begin{align}\label{detprob}
P_1(N)\approx\alpha^N \cos^{2N}\theta,
\end{align} 
which recovers the quantum Zeno effect for $\alpha=1$, $\theta=\pi/2N$ and $N\rightarrow \infty$. However, in the presence of thermal input fields,  $P_1(N)$ does  not become unity when $N\rightarrow \infty$ since $\alpha<1$.
\section{QZE in the presence of absorption}
The discussion so far considered non-absorbing beam splitters at nonzero temperatures. The only loss mechanism in that case is the reflection of  photons from any of the beam splitters. Another loss mechanism to consider is the absorption of photon by the beam splitters. The loss of photon due to either absorption or reflection is equivalent to dissipation in the system.   The corresponding non-unitary evolution of the state is simulated using the Monte Carlo wavefunction (MCWF) method [28].  This  method involves choosing unitary and non-unitary evolutions (quantum jump) of a quantum state probablistically [29]. This stochastic evolution of the quantum state forms a quantum trajectory and the ensemble average over many trajectories reproduces the actual evolution of the quantum state.

For each beam splitter in the array shown in Fig. \ref{setup}, which are assumed to be identical, the absorption, transmission or reflection of a photon is decided probabilistically. The absorption and reflection of the photon are loss mechanisms in the experimental configuration under consideration. In the language of MCWF, the absorption or reflection of the photon is considered to be a quantum jump as it collapses the output state of the beam splitter to the vacuum state.  This, in turn, means that no photon enters to the next beam splitter in the array.
 
MCWF procedure requires to define a relaxation Hamiltonian that describes the non-unitary evolution of the state. The relaxation Hamiltonian is [27]
\begin{align}\label{relax}
H_{relax}=-i\frac{\gamma}{2} \sigma_+\sigma_--i\frac{\sin^2\theta}{2} \sigma_+\sigma_-,
\end{align}
where $\sigma_+=\ket{10}\bra{00}$, $\sigma_-=\ket{00}\bra{10}$ are the quantum jump operators. Here, $\gamma$ is the absorption coefficient and $\sin^2\theta$ is the probability for reflection. In the case of identical beam splitters, $\theta=\pi/2N$. The first term in Eqn. \ref{relax} corresponds to absorption  and the second term accounts for reflection.  Now, to decide whether a photon undergoes the unitary process (transmission) or quantum jump (reflection or absorption)  in a particular beam splitter,  the quantity 
\begin{align}
\delta p_n&=i\langle \psi_{in}^{(n)}|H_{relax}-H_{relax}^\dagger |\psi_{in}^{(n)}\rangle \nonumber\\
&=(\gamma+\sin^2\theta) \bra{\psi_{in}^{(n)}}\sigma_+\sigma_-\ket{\psi_{in}^{(n)}}, 
\end{align}
is compared with a random number $r_n$ sampled from a uniform distribution. Here, $\ket{\psi_{in}^{(n)}}$ is the input state to $n$th beam splitter.  For a single-photon input state, $\ket{\psi_{in}^{(n)}} = \ket{10}$. If $r_n \ge\delta p_n$,  the photon is transmitted through the beam splitter (unitary process), and we take the probability of detecting the photon at the transmission port to be $P_1(n)=1$. If $r_n<\delta p_n$, then a quantum jump occurs, \textit{i.e.} either photon is reflected or absorbed, and we take $P_1(n)=0$. Once, a quantum jump occurs, we stop the evolution and take $P_1(n)=0$ for subsequent beam splitters. This  procedure is  followed for all the beam splitters. Obtaining the single-photon transmission probability $P_1$ for all the beam splitters completes a single trajectory. 

\begin{figure}
\begin{center}
\includegraphics[height=5cm,width=8cm]{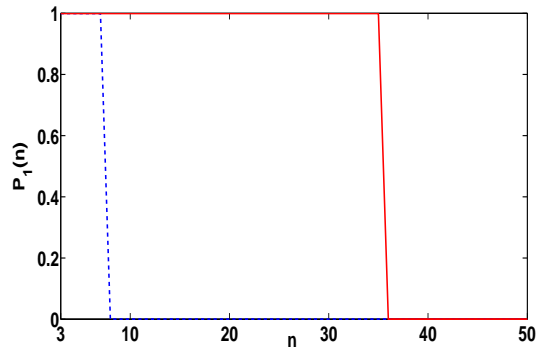}
\caption{(Color online) Two Monte Carlo trajectories for $P_1(n)$ as a function of location of the beam splitter $n$. In first realization, quantum jump occurs at 8th beam splitter and in another realization quantum jump occurs at 36th beam splitter. }
\label{trajectories}
\end{center}
\end{figure}

Fig. \ref{trajectories} shows two typical MCWF trajectories when the arrangement has fifty beam splitters.  One of the trajectories (dashed line) corresponds to a quantum jump occurring when the photon reaches the  $8$th beam splitter   and the other trajectory (continuous line) corresponds to a quantum jump occurring at the  $36$th beam splitter. The probability of detecting a photon at the transmission port becomes zero at the subsequent beam splitters after a quantum jump occurs.  As the absorption or reflection of the photon in the array configuration may occur at any one of the beam splitters, the trajectory obtained using MCWF procedure represents such a particular event (realization) of the experiment. However, this is only a possible sequence of events for the  evolution of the quantum state. Averaging over many such realizations gives the probabilities for transmission.

\begin{figure}[h!]
\centering
\includegraphics[height=5cm,width=9cm]{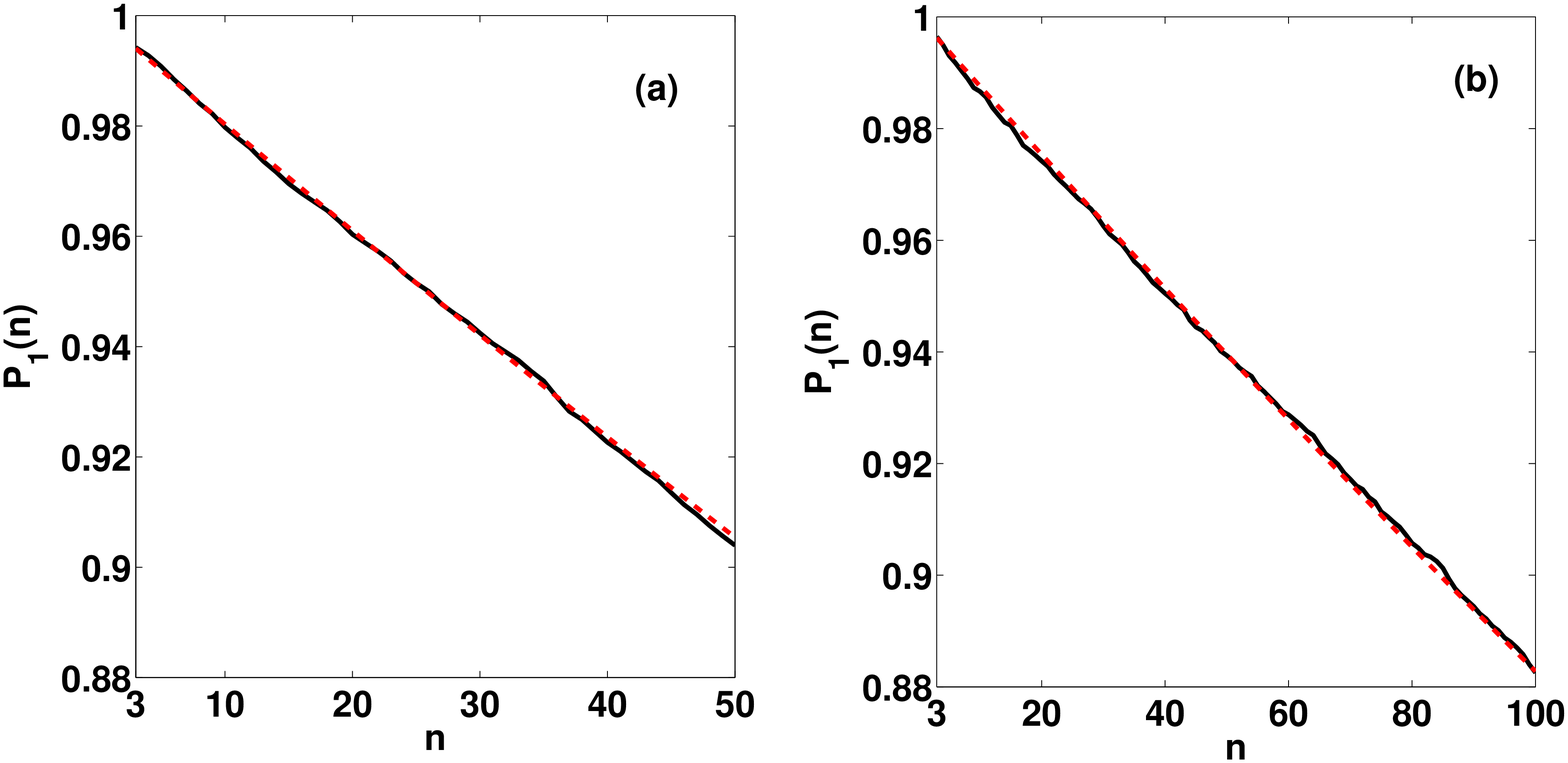}
\caption{(Color online) Single-photon detection probability as a function of $n$ for $(a)N=50$ and $(b) N=100$ in the presence of absorption. Here $\gamma=0.001$. Dashed line corresponds to the analytical result obtained from  Eqn. \ref{absqze} and solid line corresponds to numerical results from  MCWF simulation. }
\label{QZEabsorption}
\end{figure}

We calculate the probability $P_1(n)$ of detecting  photon at each beam splitter,  using the MCWF procedure and  average over 5000 realizations.  In Fig. \ref{QZEabsorption},  $P_1(n)$ is shown as a function of $n$ for two different configurations having $N=50$ and $N=100$. Here $n$ is the location of a beam splitter in the array and  $N$ is the total number of beam splitters in the array. The results obtained from MCWF method are compared with the analytical expression
\begin{align}\label{absqze}
P_1(n)=e^{-(\gamma+\sin^2\theta) n}\approx e^{-\gamma n}\left(\cos\left(\frac{\pi}{2N}\right)\right)^{2n}.
\end{align}
The above approximation is true for large $N$. For $\gamma=0$, we recover the expression given in Eqn. \ref{QZEbeamsplitter} relevant to the ideal case.   However, for $\gamma \neq 0$, the probability for detecting the photon at the port $c_N$ is zero for large $n$. As each beam splitter provides a pathway for photon loss  through absorption, it is nearly certain that photon will be lost when $n$ is large.
\begin{figure}[h!]
\centering
\includegraphics[height=5cm,width=9cm]{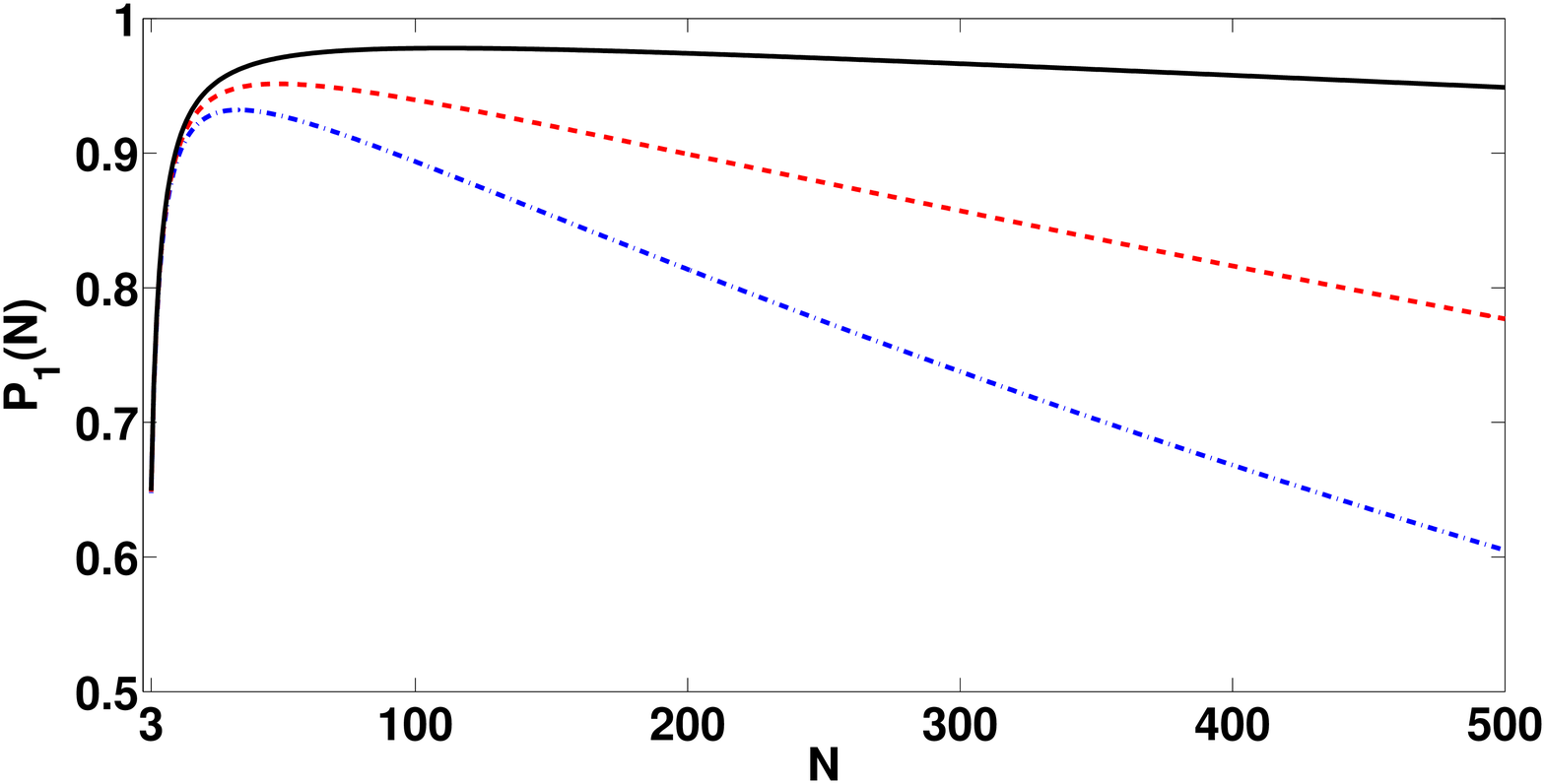}
\caption{(Color online) Single-photon detection probability $P_1(N)$ at final port in an arrangement consisting of $N$ number of beam splitters. Here $\gamma= 0.0001$ (continuous),0.0005 (dashed) and 0.001 (dot-dashed).}
\label{CriticalBeamsplitter}
\end{figure}

Now, we plot the probability of finding the photon at the end of the array, that is, $P_1(N)$ as a function of $N$ in Fig. \ref{CriticalBeamsplitter} for various values of $\gamma$. As can be seen from the figure, the probability increases to a maximum at a particular $N$ and then decreases. We denote this value of $N$ to be $N_c$, the critical number of beam splitters.  In the ideal situation wherein all the beam splitters are both ideal and identical, the experimental configuration requires a large, ideally, infinite number of beam splitters.   However, the fact that $P_1(N)$ exhibits a maximum at $N=N_c$ implies that the number of beam splitters should be limited to $N_c$.  Increasing the number beyond $N_c$ decreases the photon detection probability.   The optimal number of beam splitters $N_c$ and $\gamma$ are related through
\begin{align}\label{critical} 
\gamma=2\log\left[\cos\left(\frac{\pi}{2N_c}\right)\right]+\frac{\pi}{N_c}\tan\left(\frac{\pi}{2N_c}\right).
\end{align}
For large $N_c$, Eqn. \ref{critical} yields 
$N_c\approx\frac{1}{2}\sqrt{\frac{\pi}{\gamma}}$. 
Note that this expression is not to be interpreted to mean that with higher absorption coefficient we require fewer number of beam splitters.  Rather, the expression tells us that increasing the number of beam splitters beyond the optimal value will reduce the possibility of photon transmission.  On the lower side, the number of beam splitters is insufficient to mimic the frequent  measurement required to realize the quantum  Zeno effect.   
\section{Summary}

Theoretical analyses of a beam splitter array to realize quantum Zeno effect indicates the stringent requirement for ideal beam splitters.   Unequal reflectivities, absorption and nonzero temperature reduce the probability for single photon transmission through the array.    The array configuration is not apt for testing quantum Zeno effect as realizing identical optical components is difficult. A configuration in which the photon is routed through the same beam splitter using reflectors is a better alternative  compared to an array of beam splitters.  Simulations of the experiment using the  Monte Carlo wavefunction method indicate that there is an  optimum number of beam splitters to achieve   maximum  transmission probability.  The optimal  number of beam splitters in the array configuration or, equivalently, the number of photon traversals in the single beam splitter configuration is inversely dependent on the square root of the absorption coefficient.  Of course, the optimal number of beam splitters or photon traversals  is to closely mimic the frequent measurements and maximize the photon transmission probability in the presence of absorption required to realize the quantum Zeno effect. \\


\noindent
\textbf{Authors' contributions}\\%
 N.M and S.S formulated the problem, performed the analytic calculations and wrote the manuscript. A.R performed the numerical simulation. S.S supervised the project.\\



\begin{thebibliography}{0}

\bibitem{1} B. Misra and E. C. G. Sudarshan, {\it Journal of Mathematical Physics} \textbf{18}, 756 (1977).

\bibitem{2} P. Facchi, V. Gorini, G. Marmo, S. Pascazio, and E. Sudarshan, \textit{Physics Letters} \textbf{A 275}, 12 (2000).

\bibitem{3} T. Petrosky, S. Tasaki, and I. Prigogine, \textit{Physics Letters
A} \textbf{151}, 109 (1990).

\bibitem{4} T. Petrosky, S. Tasaki, and I. Prigogine, \textit{Physica A: Statistical Mechanics and its Applications} \textbf{170}, 306 (1991).

\bibitem{5} P. Facchi and S. Pascazio, \textit{Phys. Rev. Lett.} \textbf{89}, 080401 (2002).
\bibitem{6} W. M. Itano, D. J. Heinzen, J. J. Bollinger, and D. J.
Wineland, \textit{Phys. Rev. A} \textbf{41}, 2295 (1990).
\bibitem{7} A. G. Kofman and G. Kurizki, \textit{Phys. Rev. A} \textbf{54}, R3750
(1996).
\bibitem{8} M. C. Fischer, B. Gutierrez-Medina, and M. G. Raizen,
\textit{Phys. Rev. Lett.} \textbf{87}, 040402 (2001).
\bibitem{9} A. G. Kofman, G. Kurizki, and T. Opatrny, \textit{Phys. Rev.} \textbf{63}, 042108 (2001).
\bibitem{10} E.W. Streed, J. Mun, M. Boyd, G. K. Campbell, P. Medley, W. Ketterle, and D. E. Pritchard, \textit{Phys. Rev. Lett.}
\textbf{97}, 260402 (2006).
\bibitem{11} G. A. Alvarez, D. D. B. Rao, L. Frydman, and G. Kurizki, \textit{Phys. Rev. Lett.} \textbf{105}, 160401 (2010).
\bibitem{12} D. D. Bhaktavatsala Rao and G. Kurizki, \textit{Phys. Rev. A}
\textbf{83}, 032105 (2011).

\bibitem{13} P.-W. Chen, D.-B. Tsai, and P. Bennett, \textit{Phys. Rev. B}
\textbf{81}, 115307 (2010).
\bibitem{14} J. M. Raimond, P. Facchi, B. Peaudecerf, S. Pascazio,
C. Sayrin, I. Dotsenko, S. Gleyzes, M. Brune, and
S. Haroche, \textit{Phys. Rev. A} \textbf{86}, 032120 (2012).
\bibitem{15} F. Schfer, I. Herrera, S. Cherukattil, C. Lovecchio, F. S.
Cataliotti, F. Caruso, and A. Smerzi, \textit{Nature Communications} \textbf{5}, 3194 (2014).
\bibitem{16} D. H. Slichter, C. Mller, R. Vijay, S. J. Weber, A. Blais,
and I. Siddiqi, \textit{New Journal of Physics} \textbf{18}, 053031 (2016).
\bibitem{17} J. D. Franson, B. C. Jacobs, and T. B. Pittman, \textit{Phys.
Rev. A} \textbf{70}, 062302 (2004).
\bibitem{18} M. Das, K. Thapliyal, B. Sen, J. Perina, and A. Pathak,
\textit{Phys. Rev. A} \textbf{103}, 013713 (2021).
\bibitem{19} G. S. Agarwal and S. P. Tewari, \textit{Physics Letters A} \textbf{185},
139 (1994).
\bibitem{20} H. Salih, Z. H. Li, M. Al-Amri and M. S. Zubairy, \textit{Phys. Rev. Lett} \textbf{110},
170502 (2013).
\bibitem{21} I. Gonzalo, M. A. Porras and A. Luis, \textit{Eur. J. Phys.} \textbf{36}, 045001 (2015).
\bibitem{22} G. S. Agarwal, \textit{Quantum Optics} (Cambridge University
Press, 2013).
\bibitem{23} C. Gerry and P. Knight, \textit{Introductory Quantum Optics}
(Cambridge University Press, 2004).
\bibitem{24} D. A. B. Miller, \textit{Optica} \textbf{2}, 747 (2015).
\bibitem{25} C. M. Chandrashekar, S. Melville, and T. Busch, \textit{Journal
of Physics B: Atomic, Molecular and Optical Physics} \textbf{47},
085502 (2014).
\bibitem{26} P. Kwiat, H. Weinfurter, and A. Zeilinger, \textit{Scientific
American}\textbf{ 275}, 72 (1996).

\bibitem{27} M. Dakna, L. Knll, and D.-G. Welsch, \textit{Optics Communications} \textbf{145}, 309 (1998).
\bibitem{28} K. Molmer, Y. Castin, and J. Dalibard, \textit{J. Opt. Soc. Am.
B} \textbf{10}, 524 (1993).
\bibitem{29} J. Dalibard, Y. Castin, and K. Molmer, \textit{Phys. Rev. Lett.}
\textbf{68}, 580 (1992).
\end{thebibliography}
\end{document}